\documentclass{amsart}
\usepackage{amssymb}
\usepackage{amsmath}
\usepackage{amsfonts}

\newtheorem{theorem}{Theorem}
\theoremstyle{plain}

\newtheorem{lemma}{Lemma}

\numberwithin{equation}{section}
\input{tcilatex}

\begin{document}
\title[A nonlinear evolution equation related to nonstationary Dirac-type
system]{A nonlinear evolution equation with 2 + 1 dimensions related to
nonstationary Dirac-type system}
\author{Mansur I Ismailov}
\address{Department of Mathematics, Gebze Institute of Technology,
Gebze-Kocaeli \ 41400, Turkey}
\email{mismailov@gyte.edu.tr }
\date{15, August, 2009}
\dedicatory{}

\begin{abstract}
In this paper the inverse scattering problem for the nonstationary
Dirac-type system on the whole plane was considered. A nonlinear evolution
sytem of equation related to nonstationary Dirac-type system is introduced
and the solviblity of this sytem using the IST method is studied.
\end{abstract}

\maketitle

\section{Introduction}

As is known, every linear scattering problem determines a class of nonlinear
evolution equations for which the inverse scattering transform (IST) method
is suitable to integrate these equations.\textrm{\ }The investigation of the
nonlinear evolution equations by the IST method has been stated in [1, 2].
The essence of this method is to represent this nonlinear equation as a Lax
equation

\begin{equation}
\frac{\partial \mathbf{L}}{\partial t}=\left[ \mathbf{L},\mathbf{A}\right] 
\tag{0.1}
\end{equation}%
by using the Lax pair$\ \mathbf{L}$ and $\mathbf{A}$. The equation in the
form of (0.1) allows to investigate evolution of scattering data instead of
the evolution of the operator $\mathbf{L}$, if the inverse scattering
problem is investigated for the operator $\mathbf{L}$.

The IST method for the nonlinear evolution equations with 1+1 dimensions
(one space and one time dimensions) have been described in a variety reviews
and monographs. The generalization of the IST method to nonlinear evolution
equations with 2+1 dimensions (two space and one time dimension ) has been
developed in monograph [3]. In this direction one can also notice the papers
[4 - 7], in which the inverse scattering problem for the first order systems
is studied and this inverse problem was applied to integration of $N$- wave
interactions, Davey-Stewartson and Kodomtsev-Petviashvili equations.

Let $\mathbf{L}=\frac{\partial }{\partial y}-\mathbf{M}$, where $\mathbf{M}$
is an ordinary differential operator in $x$. Then equation (0.1) is
rewritten as (see [2])

\begin{equation}
\left[ \frac{\partial }{\partial y}-\mathbf{M},\frac{\partial }{\partial t}-%
\mathbf{A}\right] =0  \tag{0.2}
\end{equation}%
and it is a relation for ordinary differential operators. Some nonlinear
evolution equations in 2+1 dimension are presented in [8] by using
commutativity condition (0.2) when $\mathbf{M}$ is an scalar coefficients
ordinary differential operator in $x$.

This article is organized as follows: In Section 2, we consider the inverse
scattering problems for the $\mathbf{L}=\frac{\partial }{\partial y}-\mathbf{%
M}$, with $\mathbf{M}=\sigma \frac{\partial }{\partial x}+Q(x,y)$ in the
whole plane. We introduce the scattering data for the operator $\mathbf{L}$
in the plane (the minimal information for a unique restoration of
matrix-function $Q(x,y)$). All the results in Section 2 are the matrix
generalization of the L. P. Nizhnik's results for two component
nonstationary Dirac system on the whole plane in [9] (see [4] also). Taking
into account that this generalization is not difficult, so we do not give
the proofs. In Section 3, we introduce some system of nonlinear evolution
equations with 2+1 dimensions related to nonstationary Dirac-type systems
and we study the solvability of this system using the IST method. In [10],
the IST method is applied to integrate the two component nonlinear Schr\"{o}%
dinger equation, using the inverse scattering problem for stationary
Dirac-type system in whole line.

\section{\textbf{Inverse scattering problem for the Dirac-type system in the
plane}}

Let us consider a system of first-order partial differential equations (PDEs)%
\begin{equation}
\mathbf{L(}\psi )\equiv \frac{\partial \psi }{\partial y}-\sigma \frac{%
\partial \psi }{\partial x}-Q\left( x,y\right) \psi =0,\text{ }-\infty
<x,y<+\infty ,  \tag{1.1}
\end{equation}%
where

\begin{equation*}
\sigma =\left[ 
\begin{array}{ccc}
1 & 0 & 0 \\ 
0 & 1 & 0 \\ 
0 & 0 & -1%
\end{array}%
\right] ,Q=\left[ 
\begin{array}{ccc}
0 & 0 & q_{1} \\ 
0 & 0 & q_{2} \\ 
q_{3} & q_{4} & 0%
\end{array}%
\right] .
\end{equation*}%
with the measurable complex-valued functions $q_{i}$ ($i=1,\ldots ,4)$.

We give fundamental results on the inverse scattering problem for the system
(1.1) in the case where the coefficients $q_{i}($ $i=1,\ldots ,4)$ decrease
quite fast with respect to variables $x$ and $y$ at infinity. Each bounded
solution of the system (1.1) has the asymptotic

\begin{eqnarray*}
\psi \left( x,y\right) &=&\mathbf{\digamma }_{x}a_{-}\left( y\right) +o(1),\
\ y\longrightarrow -\infty \ , \\
\psi \left( x,t\right) &=&\digamma _{x}a_{+}\left( t\right) +o(1),\ \
y\longrightarrow +\infty \ .
\end{eqnarray*}%
where $\mathbf{\digamma }_{x}$ denotes the diagonal matrix shift operator,
such that for a vector function $a\left( y\right) =\func{col}\left[
a_{1}\left( y\right) ,a_{2}\left( y\right) ,a_{3}\left( y\right) \right] ,$ $%
\ \digamma _{x}a\left( y\right) =\func{col}\left[ a_{1}\left( t+x\right)
,a_{2}\left( t+x\right) ,a_{3}\left( t-x\right) \right] $

The scattering operator $S$ is defined by the equality

\begin{equation}
a_{+}\left( y\right) =\mathbf{S}a_{-}\left( y\right) .  \tag{1.2}
\end{equation}

This operator has the inverse $S^{-1}$ and $S=I+F$ and $S^{-1}=I+G,$ where $%
F=\left[ F_{ij}\right] _{i,j=1}^{3}$ and $G=\left[ G_{ij}\right]
_{i,j=1}^{3} $ are matrix integral operators. To regenerate the system
(1.1), i.e., the coefficients $q_{i}($ $i=1,\ldots ,4)$ it is sufficient to
know $F_{13},F_{23},G_{31}$ and $G_{32}$. The collection $\left\{
F_{13},F_{23},G_{31},G_{32}\right\} $ is called the scattering data for the
system (1.1) in the plane. The solution of the inverse problem reduces to
the integral equation

\begin{eqnarray}
A\left( x,y,\tau \right) -\int_{-\infty }^{y}\left[ \int_{y}^{+\infty
}A\left( x,y,z\right) \tilde{G}\left( z-x,s+x\right) dz\right] \tilde{F}%
\left( s+x,\tau -x\right) ds &=&\tilde{F}\left( y+x,\tau -x\right) ,  \notag
\\
&&  \TCItag{1.3} \\
B\left( x,y,\tau \right) -\int_{y}^{+\infty }\left[ \int_{-\infty
}^{y}B\left( x,y,z\right) \tilde{F}\left( z+x,s-x\right) dz\right] \tilde{G}%
\left( s+x,\tau -x\right) ds &=&\tilde{G}\left( y-x,\tau +x\right) ,  \notag
\end{eqnarray}%
where $\tilde{F}=\left[ 
\begin{array}{c}
F_{13} \\ 
F_{23}%
\end{array}%
\right] $ and $\tilde{G}=\left[ 
\begin{array}{cc}
G_{31} & G_{32}%
\end{array}%
\right] .$

The coefficients $q_{i}($ $i=1,\ldots ,4)$ of the system (1.1) are expressed
in terms of the solution of (1.3) by means of the equations

\begin{equation*}
\left[ 
\begin{array}{c}
q_{1} \\ 
q_{2}%
\end{array}%
\right] \left( x,y\right) =-2A\left( x,y,y\right) ,\ \left[ 
\begin{array}{cc}
q_{3} & q_{4}%
\end{array}%
\right] \left( x,y\right) =-2B\left( x,y,y\right) .
\end{equation*}

\section{Nonlinear evolution equation related to nonstationary Dirac-type
systems}

Let $\mathbf{M}$ and $\mathbf{A}$ be first order matrix operators in (0.2):%
\begin{equation*}
\mathbf{M}=\sigma \frac{\partial }{\partial x}+Q,\text{ \ }\mathbf{A}=\tau 
\frac{\partial }{\partial x}+P.
\end{equation*}

Then the equation (0.2) becomes to the form 
\begin{equation}
\left[ \frac{\partial }{\partial y}-\sigma \frac{\partial }{\partial x}-Q,%
\frac{\partial }{\partial t}-\tau \frac{\partial }{\partial x}-P\right] =0. 
\tag{2.1}
\end{equation}

Here $\tau $ and $P$ are third order square matrices. Let the matrix $\tau $
be real and diagonal: $\tau =diag(b_{1},b_{2},b_{3}),$ $b_{1}>b_{2}>b_{3}$
and the matrices $Q$ and $P$ obey the relation: $\left[ \sigma ,P\right] =%
\left[ \tau ,Q\right] $.

Take $P=\left[ 
\begin{array}{ccc}
0 & v_{12} & \frac{b_{1}-b_{3}}{2}q_{1} \\ 
v_{21} & 0 & \frac{b_{2}-b_{3}}{2}q_{2} \\ 
\frac{b_{1}-b_{3}}{2}q_{3} & \frac{b_{2}-b_{3}}{2}q_{4} & 0%
\end{array}%
\right] ,$ where the functions $v_{12}$ and $v_{21}$ are solutions of the
equations 
\begin{eqnarray*}
\frac{\partial }{\partial y}v_{12}-\frac{\partial }{\partial x}v_{12} &=&-%
\frac{b_{1}-b_{2}}{2}q_{1}q_{4}, \\
\frac{\partial }{\partial y}v_{21}-\frac{\partial }{\partial x}v_{21} &=&-%
\frac{b_{2}-b_{1}}{2}q_{2}q_{3},
\end{eqnarray*}%
respectively.

Then, the conditions (2.1) are reduced to the system of equations%
\begin{eqnarray}
\frac{\partial }{\partial t}q_{1}+k_{1}\frac{\partial }{\partial y}%
q_{1}+k_{2}\frac{\partial }{\partial x}q_{1} &=&v_{12}q_{2},  \notag \\
\frac{\partial }{\partial t}q_{2}+k_{1}\frac{\partial }{\partial y}%
q_{2}+k_{2}\frac{\partial }{\partial x}q_{2} &=&v_{21}q_{1},  \notag \\
&&  \TCItag{2.2} \\
\frac{\partial }{\partial t}q_{3}+k_{3}\frac{\partial }{\partial y}%
q_{3}+k_{4}\frac{\partial }{\partial x}q_{3} &=&-v_{21}q_{4},  \notag \\
\frac{\partial }{\partial t}q_{4}+k_{3}\frac{\partial }{\partial y}%
q_{4}+k_{4}\frac{\partial }{\partial x}q_{4} &=&-v_{12}q_{3},  \notag
\end{eqnarray}%
where $k_{1}=-\frac{b_{1}-b_{3}}{2},$ $k_{2}=-\frac{b_{1}+b_{3}}{2},k_{3}=-%
\frac{b_{2}-b_{3}}{2},k_{4}=-\frac{b_{2}+b_{3}}{2}.$

Let us denote $\mathbf{P}=\frac{\partial }{\partial t}-\tau \frac{\partial }{%
\partial y}-P$. The Lax form (2.1) of the system of equations (2.2) enables
us to apply to IST\ method for the integration.\ 

\begin{lemma}
Let $\psi $ be a solution of the Dirac-type system (1.1), whose the
coefficients $q_{i},i=1,\ldots ,4$ satisfy Eqs. (2.2). Then the function $%
\varphi =\mathbf{P}\psi $ also satisfy the system (1.1).

\begin{proof}
Let us apply the operator equation to $\psi $:%
\begin{equation*}
(\mathbf{LP}-\mathbf{PL)}\psi =\mathbf{L(P\psi )-P(\mathbf{L}\psi )=0.}
\end{equation*}

Taking into account that $\mathbf{\mathbf{L}\psi =0}$, then 
\begin{equation*}
\mathbf{L(P\psi )=0}
\end{equation*}

follows from the last equation. It means that $\mathbf{P\psi }$ is solution
of system (1.1).
\end{proof}
\end{lemma}

The next theorem is true for the evolution of the scattering data.

\begin{lemma}
Let the coefficients $q_{i},i=1,\ldots ,4$ of the Dirac-type system (1.1)
depend on $t$ as a parameter and satisfy the system of equation (2.2).
Besides that 
\begin{equation}
v_{12}(x,\pm \infty )=0,\text{ \ }v_{21}(x,\pm \infty )=0.  \tag{2.3}
\end{equation}

Then the kernels $F_{13}(y,\tau ;t),F_{23}(y,\tau ;t),G_{31}(y,\tau
;t),G_{32}(y,\tau ;t)$ of the integral operators $%
F_{13},F_{23},G_{31},G_{32} $ corresponding to the scattering operator $%
\mathbf{S}$ for the system (1.1) in the plane satisfy the system of
equations \ 
\begin{eqnarray}
\frac{\partial }{\partial t}F_{13}\left( y,\tau ;t\right) -b_{1}\frac{%
\partial }{\partial y}F_{13}\left( y,\tau ;t\right) +b_{3}\frac{\partial }{%
\partial \tau }F_{13}\left( y,\tau ;t\right) &=&0,  \notag \\
&&  \TCItag{2.4} \\
\frac{\partial }{\partial t}F_{23}\left( y,\tau ;t\right) -b_{2}\frac{%
\partial }{\partial y}F_{23}\left( y,\tau ;t\right) +b_{3}\frac{\partial }{%
\partial \tau }F_{23}\left( y,\tau ;t\right) &=&0  \notag
\end{eqnarray}

and%
\begin{eqnarray}
\frac{\partial }{\partial t}G_{31}\left( y,\tau ;t\right) +b_{3}\frac{%
\partial }{\partial y}G_{31}\left( y,\tau ;t\right) -b_{1}\frac{\partial }{%
\partial \tau }G_{31}\left( y,\tau ;t\right) &=&0,  \notag \\
&&  \TCItag{2.5} \\
\frac{\partial }{\partial t}G_{32}\left( y,\tau ;t\right) +b_{3}\frac{%
\partial }{\partial y}G_{32}\left( y,\tau ;t\right) -b_{2}\frac{\partial }{%
\partial \tau }G_{32}\left( y,\tau ;t\right) &=&0.  \notag
\end{eqnarray}

\begin{proof}
By virtue of definition of the scattering operator $\mathbf{S}$, from Lemma
1 we get%
\begin{equation}
\varphi _{+}=\mathbf{S}\varphi _{-}  \tag{2.6}
\end{equation}

where $\varphi _{\pm }=\mathbf{P}a_{\pm }$, $\mathbf{P}=\frac{\partial }{%
\partial t}-\tau \sigma \frac{\partial }{\partial y}$.

Since $a_{+}=\mathbf{S}a_{-}$ (see (1.2)), the equality 
\begin{equation}
\mathbf{PS=SP}  \tag{2.7}
\end{equation}

follows from (2.6).

Analogously, 
\begin{equation}
\mathbf{PS}^{-1}\mathbf{=S}^{-1}\mathbf{P.}  \tag{2.8}
\end{equation}

From the matrix operator equation (2.7) it follows that the kernels of the
integral operators $F_{13}$ and $F_{23}$ satisfy system of equation (2.4).
The similarly system of equation (2.5) for the kernels of the integral
operators $G_{31}$ and $G_{32}$ follows from the matrix operator equation
(2.8).
\end{proof}
\end{lemma}

According to Chapter 2 and Lemma 2, let us give a procedure for the solution
of the system (2.2) by IST method.

\begin{theorem}
The system of equations (2.2) admits integration in the class of decreasing
functions that satisfy the condition (2.3) by the IST method. The solution $%
q_{i},i=1,\ldots ,4$ of the system (2.2) is determined by formulae%
\begin{equation*}
\left[ 
\begin{array}{c}
q_{1} \\ 
q_{2}%
\end{array}%
\right] (x,y;t)=-2A(x,y,y;t),\text{ \ }\left[ 
\begin{array}{cc}
q_{3} & q_{4}%
\end{array}%
\right] (x,y;t)=-2B(x,y,y;t),
\end{equation*}

where the vector functions $A$ and $B$ are the solutions of the integral
equations%
\begin{eqnarray*}
&&%
\begin{array}{c}
A\left( x,y,\tau ;t\right) -\int_{-\infty }^{y}\left[ \int_{y}^{+\infty
}A\left( x,y,z;t\right) \tilde{G}\left( z-x,s+x;t\right) dz\right] \tilde{F}%
\left( s+x,\tau -x;t\right) ds \\ 
=\tilde{F}\left( y+x,\tau -x;t\right) ,%
\end{array}
\\
&&%
\begin{array}{c}
B\left( x,y,\tau ;t\right) -\int_{y}^{+\infty }\left[ \int_{-\infty
}^{y}B\left( x,y,z;t\right) \tilde{F}\left( z+x,s-x;t\right) dz\right] 
\tilde{G}\left( s+x,\tau -x;t\right) ds \\ 
=\tilde{G}\left( y-x,\tau +x;t\right) ,%
\end{array}%
\end{eqnarray*}%
where $\tilde{F}=\left[ 
\begin{array}{c}
F_{13} \\ 
F_{23}%
\end{array}%
\right] $ and $\tilde{G}=\left[ 
\begin{array}{cc}
G_{31} & G_{32}%
\end{array}%
\right] $ are satisfied the evolution equations (2.4) and (2.5).
\end{theorem}

Thus the system of nonlinear evolution equation (2.2) admits the integration
by IST method. Let us denote by $\mathbf{\Pi }$\textbf{\ }the operator
transforming the coefficients $q(t)=\func{col}%
(q_{1}(x,y;t),q_{2}(x,y;t),q_{3}(x,y;t),q_{4}(x,y;t))$ of the equation (1.1)
to scattering data $T(t)=\func{col}(F_{13}(y,\tau ;t),F_{23}(y,\tau
;t),G_{31}(y,\tau ;t),G_{32}(y,\tau ;t))$ as follows:%
\begin{equation*}
\mathbf{\Pi }:q(t)\longmapsto T(t).
\end{equation*}

Then the solution of the system (2.2) can be represent as%
\begin{equation}
q(t)=\mathbf{\Pi }^{-1}e^{t\mathbf{M}}T  \tag{2.9}
\end{equation}%
where $T=T(0)$ i.e. $T=\func{col}(F_{13}(y,\tau ),F_{23}(y,\tau
),G_{31}(y,\tau ),G_{32}(y,\tau ))$ and 
\begin{equation*}
\mathbf{M}=\left[ 
\begin{array}{cccc}
b_{1}\frac{\partial }{\partial y}-b_{3}\frac{\partial }{\partial \tau } & 0
& 0 & 0 \\ 
0 & b_{2}\frac{\partial }{\partial y}-b_{3}\frac{\partial }{\partial \tau }
& 0 & 0 \\ 
0 & 0 & b_{1}\frac{\partial }{\partial \tau }-b_{3}\frac{\partial }{\partial
y} & 0 \\ 
0 & 0 & 0 & b_{2}\frac{\partial }{\partial \tau }-b_{3}\frac{\partial }{%
\partial y}%
\end{array}%
\right] .
\end{equation*}

The formula (2.9) is a specified form of solution $q(t)$ of (2.2).


\begin{thebibliography}{99}
\bibitem{[1]} M. J. Ablowitz and H. Segur, \textit{Solitons and the inverse
scattering transform.} SIAM Stud. Appl. Math. \textbf{4}, Philadelphia, 1981.

\bibitem{[2]} V. E. Zakharov, S. V. Manakov, S. P. Novikov, and L. P.
Pitaievski, \textit{Theory of solitons. The inverse scattering method.}
Nauka, Moscow, 1980 (in Russian); English transl. Consultants bureau (New
York: Plenum) 1984.

\bibitem{[3]} B. G. Konopelchenko, \textit{Introduction to multidimensional
integrable equations. The inverse spectral transform in 2+1 dimensions}.
Plenum Press, New York, 1992. 292 pp.

\bibitem{[4]} L. P. Nizhnik, \textit{The inverse scattering problem for
hyperbolic equations and their application to nonlinear integrable systems.}
Reports on Math \ Phys. \textbf{26} (2), 261-283 (1988).

\bibitem{[5]} L. P. Nizhnik and M. D. Pachinaiko, \textit{Integration of the
nonlinear two-dimensional spatial Schr\"{o}dingerequation by the
inverse-problem method, }Functional Anal. Appl., 1982, \textbf{16 }(1), 66-69

\bibitem{[6]} A. S. Fokas and L. Y. Sung, \textit{On the solvability of the }%
$N$\textit{- Wave, Davey-Stewartson and Kodomtsev-Petviashvili equations,}\
Inverse Problems, 1992, \textbf{8}, 673-708.

\bibitem{[7]} A. S. Fokas and M. J. Ablowits, \textit{On the inverse
scattering transform of multidimensional nonlinear equations related to
first-order systems in the plane}, J. Math. Phys., 1984,\textbf{\ 25 }(8),
2494-2505.

\bibitem{[8]} B. G. Konopelchenko and V. G. Dubrovsky, \textit{Some new
integrable nonlinear evolution equations in 2+1 dimensions}. Phys. Lett. A
102 (1984), no. 1-2, 15--17

\bibitem{[9]} L. P. Nizhnik, \textit{Inverse nonstationary scattering
problems.} Nauk. Dumka, Kiev,1973 (in Russian).

\bibitem{[10]} S. V. Manakov, \textit{On the theory of two-dimensional
stationary self-focusing of electromagnetig waves},\ Sov. Phys. JETP 38,
248-253 (1974).
\end{thebibliography}
\end{document}